\documentstyle[aps,twocolumn]{revtex}
\input epsf

\newcommand{\tnp}{$T_{\rm N}(P)$~}

\newcommand{\crg}{CeRu$_{2}$Ge$_2$~}

\newcommand{\cps}{CePd$_{2}$Si$_2$~}
\newcommand{\cng}{CeNi$_{2}$Ge$_2$~}

\newcommand{\ccs}{CeCu$_{6}$~}
\newcommand{\ccfa}{CeCu$_{5}$Au~}
\newcommand{\cca}{CeCu$_{6-x}$Au$_x$~}

\newcommand{\etal}{{\it et al.}~}

\begin{document}

\title{From an antiferromagnet to a heavy-fermion system: CeCu$_5$Au under pressure}
\author{H. Wilhelm, S. Raymond, D. Jaccard, O. Stockert$^{\dagger}$, 
            and H. v. L\"ohneysen$^{\dagger}$}
\address{D\'epartement de Physique de la Mati\`ere Condens\'ee, Universit\'e de Gen\`eve, 
         Quai Ernest-Ansermet 24, 1211 Geneva 4, Switzerland\\ 
 $^{\dagger}$ Physikalisches Institut, Universit\"at 
               Karlsruhe, 76128 Karlsruhe, Germany}
%\date{\today}
\maketitle
\widetext
\begin{abstract}
The electrical resistivity $\rho(T)$ of single crystalline 
CeCu$_5$Au under pressure was measured in the temperature range 
30~mK$<T<300$~K. Pressure suppresses the antiferromagnetic order 
($T_{\rm N}=2.35$~K at ambient pressure) and drives the system into 
a non--magnetic heavy--fermion state above $P_{\rm c}=4.1(3)$~GPa. 
The electrical resistivity shows a deviation from a $T^2$ dependence 
of a Fermi--liquid in the pressure range 1.8~GPa~$\leq P\leq 5.15$~GPa. The 
$\rho(T)$--curves can be compared with those of \cca at different 
Au concentrations. Just before the long--range magnetic order 
vanishes, a possibly superconducting phase (at $T_{\rm c}=0.1$~K and 
$P=3.84$~GPa) occurs, pointing to a coexistence of antiferromagnetic 
order and superconductivity. This new phase is only seen in a narrow 
pressure interval $\Delta P=0.4$~GPa.  
\end{abstract}
\vspace*{0.5cm}
\noindent
$[$phase transition, high pressure, electrical resistivity, non--Fermi--liquid$]$
\vspace*{0.7cm}
\narrowtext

\tighten
%###########################################################################
%                                                                          #
%                      Introduction                                        #
%                                                                          #
%###########################################################################

\begin{flushleft}
{\bf 1. Introduction}
\end{flushleft}
In many heavy--fermion (HF) metals antiferromagnetic (AFM) quantum 
critical phenomena have been observed. Ternary Ce--based compounds 
like \cps \cite{MATHU98}, \crg \cite{WILHE99}, and \cca 
\cite{LOEHN94} can be tuned to a quantum critical point (QCP) either 
by pressure or doping. According to the spin--fluctuation theory 
\cite{HERTZ76,MILLI93,MORIY95} the temperature dependence of the 
electrical resistivity $\Delta\rho(T)\propto T^n$ shows a deviation 
from the conventional Fermi--liquid (FL) behaviour ($n=2$) in a 
limited temperature range. Furthermore, the N\a'eel temperature 
$T_{\rm N}$ approaches zero in a characteristic way: $T_{\rm 
N}\propto (g_{\rm c} - g)^m\rightarrow 0$, where $g$ and $g_{\rm c}$ 
are the tuning parameter (like concentration $x$ or pressure $P$) and 
its critical value. For AFM fluctuations the exponent is 
$m=2/3$. This prediction has not been found unequivocally in 
experiment. Recent high pressure experiments on \crg \cite{WILHE99} 
gave $m=0.70\pm 0.08$ but reported variations on other compounds are 
mainly linear in $x$ or $P$ (for more details see 
Ref.~\cite{HUXLE96} and references therein). 

The anomalous exponents found for the temperature dependence of 
$\rho(T)$ of \cps and \cng ($n=1.2$ \cite{MATHU98} and 1.37 
\cite{GEGEN99}, respectively) as well as the logarithmic temperature 
dependence of the specific heat $C/T\propto \ln(T^{\star}/T)$ 
observed in \cca (see e.~g.~Ref.~\cite{LOEHN98}) point to a 
fundamental breakdown of FL theory \cite{COLEM99}. The results on 
high quality samples of \cps \cite{MATHU98} and \cng \cite{GEGEN99} 
have shown that such a non--Fermi--liquid (NFL) behaviour occurs in 
a small temperature interval. Furthermore, close to the QCP 
superconductivity emerges in some cases. On the other hand, it is 
not clear if disorder will modify or even produce the various NFL 
properties. In a recent theoretical explanation \cite{ROSCH99} it 
was argued that the anomalies in $\rho(T)$ can be attributed to the 
interplay between quantum critical AFM fluctuations and impurity 
scattering in a conventional FL. Such a consideration, however, is 
not relevant 
\pagebreak
\vspace*{4.5cm}
\noindent
for the $\Delta\rho(T)\propto T$ behaviour, observed in 
CeCu$_{5.9}$Au$_{0.1}$ \cite{LOEHN94}, where the anisotropy of the 
spin fluctuations seems to play an important role 
\cite{ROSCH97,STOCK98}. 

The stoichiometric \ccfa compound has the highest $T_{\rm N}=2.35$~K 
of all \cca alloys with $0.1\leq x\leq 1.0$ \cite{PIETR95} where 
$T_{\rm N}$ varies linearly with $x$. At the critical concentration 
$x_c=0.1$ the system loses its long--range magnetic order and the 
Kondo effect dominates the RKKY interaction ($T_{\rm K}=6.2$~K in 
CeCu$_6$ \cite{SCHLA93}). At very low temperature ($T<0.3$~K) a FL 
behaviour is found in \ccs \cite{STEWA84}. The $T_{\rm N}(x)$ 
variation seems to be related to the increase of the unit--cell 
volume upon doping, but changes in the band structure have to be 
considered if the different $T_{\rm N}(x)$ and $T_{\rm N}(P)$ 
dependence at equal volume have to be explained. Since in the alloys 
a certain disorder and a structural transition at low temperature 
(for $x<0.15$) exist, measurements on the stoichiometric and single 
crystalline \ccfa \cite{STOCK99} offer a unique possibility to study 
the pure pressure or volume effect on the magnetic ordering 
temperature. 

Here we present electrical resistivity measurements on single 
crystalline \ccfa under high pressure ($P<8$~GPa). The four--point 
resistance was measured on a sample in a clamped high pressure cell 
which was cooled down in a dilution refrigerator ($T>30$~mK). 
Details of the high pressure set--up can be found elsewhere 
\cite{WILHE99}. 

%###########################################################################
%                                                                          #
%                      Experimental Details and Results                    #
%                                                                          #
%###########################################################################

\begin{flushleft}
{\bf 2. Results and Discussion}
\end{flushleft}

\begin{figure}
\epsfxsize=100mm \epsfbox{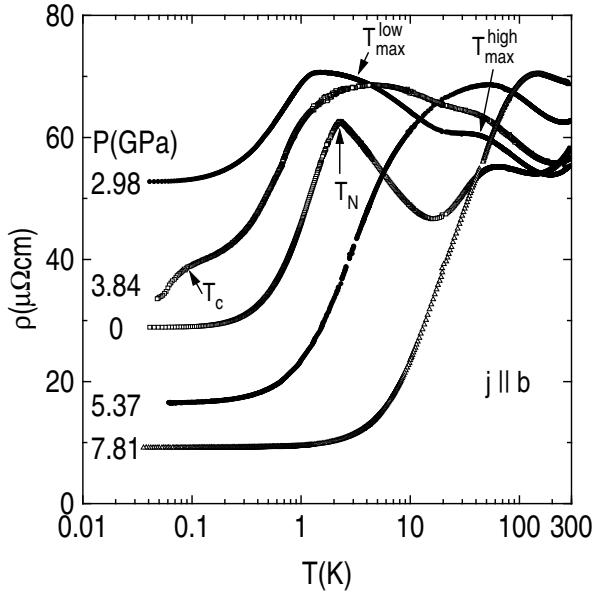}
\vspace*{-6cm}
\caption{Temperature dependence of the electrical resistivity of \ccfa
at selected pressures. The curve at $P=3.84$~GPa shows a sudden drop
at $T=0.1$~K, pointing to a new low temperature phase which exists
only in a small pressure range ($\Delta P=0.4$~GPa).}
\label{RhoVsT}
\end{figure}

Representative electrical resistivity curves $\rho(T)$ of \ccfa are 
shown in a semi--logarithmic plot in Fig.~\ref{RhoVsT}. The low 
temperature part of the ambient pressure curve is identical to that 
reported in Ref.~\cite{LOEHN98}. Below $T_{\rm N}=2.35$~K the 
antiferromagnetically ordered phase is entered, clearly visible by 
the pronounced cusp in $\rho(T)$. A negative logarithmic slope is 
present above $T_{\rm N}$ up to 10~K, reflecting the presence of the 
Kondo effect ($T_{\rm K}=1.8$~K \cite{PASCH94}). A maximum in 
$\rho(T)$ develops at higher temperature ($T_{\rm max}^{\rm 
high}\approx 60$~K). It becomes less pronounced and shifts slightly 
down in temperature as pressure is applied. At a moderate pressure 
of $P=2.98$~GPa a second maximum at $T_{\rm max}^{\rm low}=3.5$~K 
appears, which is well separated from the entrance into the AFM 
state at $T_{\rm N}=1.4$~K. It might be related to an already 
enhanced Kondo effect and could point to the development of a 
coherent state. The $\rho(T)$--curve at $P=3.84$~GPa shows a 
peculiar low temperature behaviour. The entrance into a magnetically 
ordered phase at $T_{\rm N}\approx 1$~K is still visible but at 
$T=0.1$~K the resistivity drops suddenly by more than 10\%, indicating 
the occurrence of a new phase. Its possible nature will be discussed 
below. At this pressure the two maxima in $\rho(T)$ are still 
present whereas at higher pressure only $T_{\rm max}^{\rm high}$ 
remains and a FL behaviour is observed at low temperature. We
mention that the residual resistivity exhibits a strong pressure 
dependence ($cf$. Fig.~\ref{RhoVsT}) and details will be given 
elsewhere \cite{WILHE99d}. 

%############################################################################
%                                                                           #
%                   FIGURE 1                                                #
%                                                                           #
%############################################################################

\begin{figure}
\epsfxsize=100mm \epsfbox{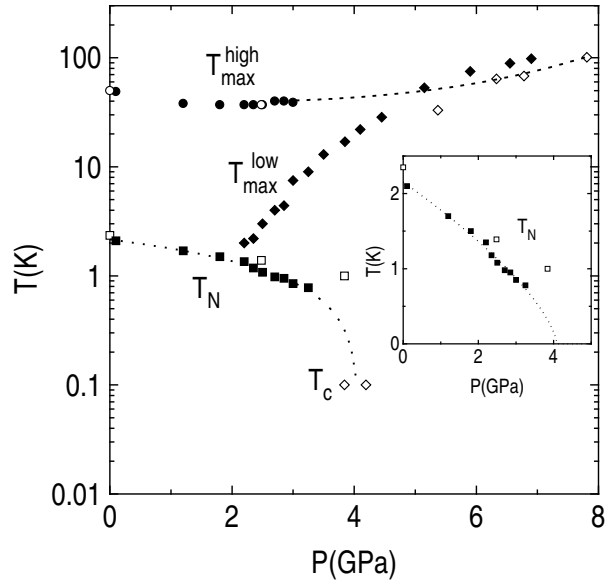}
\vspace*{-6cm}
\caption{Pressure dependence of the characteristic temperatures in $\rho(T)$
of CeCu$_5$Au in a semilogarithmic plot. The N\'eel temperature 
($T_{\rm N}$) scales to zero at $P_{\rm c}=4.1(3)$~GPa. Two maxima 
in $\rho(T)$ at $T_{\rm max}^{\rm low}$ and $T_{\rm max}^{\rm 
high}$, related to the Kondo effect, seem to merge above 4~GPa. In 
one experiment (open symbols) indications of a new phase (possibly 
superconducting) below $T_{\rm c}=0.1$~K ($\Diamond$) was found. In 
the inset the $T_{\rm N}(P)$ dependence is shown in a linear plot.} 
\label{tallccfa}
\end{figure}

In Fig.~\ref{tallccfa} the pressure dependence of the characteristic 
temperatures in $\rho(T)$ of CeCu$_5$Au, obtained on two different 
pieces cut from the same single crystal, are shown. The open squares 
show that in one experiment the magnetic order was still 
visible at $P=3.84$~GPa, where a pronounced drop in $\rho(T)$ 
occurred at $T_{\rm c}= 0.1$~K ($\Diamond$). Traces of this drop 
were also found at 4.19~GPa. In a second experiment (filled symbols in 
Fig.~\ref{tallccfa}) no signs of magnetism above 3.2~GPa and of a 
drop in resistivity around 3.8~GPa have been found. The 
\tnp dependence scales to zero like $T_{\rm N} \propto (P_{\rm c} - 
P)^m$ at a critical pressure $P_{\rm c}=4.1~\pm 0.3$~GPa with an 
exponent $m=0.68~\pm 0.11$ (inset Fig.~\ref{tallccfa}). This value is in 
good agreement with $m=2/3$, predicted within the spin fluctuation 
theory \cite{MORIY95}. 

%############################################################################
%                                                                           #
%                   FIGURE 2                                                #
%                                                                           #
%############################################################################

The Kondo temperature in \ccfa at low pressure is small in 
comparison to the crystal field (CF) splitting $\Delta_{\rm 
CF}^{(1)} \approx 100$~K and $\Delta_{\rm CF}^{(2)}\approx 160$~K 
\cite{STROK93}. Therefore, as often observed in this situation for 
other compounds, the magnetic resistivity has two maxima at $T_{\rm 
max}^{\rm low}$ and $T_{\rm max}^{\rm high}$, whose high temperature 
sides are a sign of the Kondo scattering on the ground state and 
excited CF levels, respectively, and whose low temperature sides 
reflect the onset of a coherent heavy--fermion state and the freezing 
of scattering from CF levels. The pressure variation of $T_{\rm 
max}^{\rm low}$ might point to the possibility of an enhanced 
screening (induced by pressure) of the magnetic moments by the 
conduction electrons and thus to an strengthened role of the Kondo 
effect. Consequently, the anomaly at $T_{\rm max}^{\rm low}$ has to 
be related to $T_{\rm K}$. Both anomalies in $\rho(T)$ seem to merge 
above 4~GPa, indicating the entrance into an intermediate valence 
regime where the Kondo temperature becomes of the order of the CF 
splitting. 

In the case of two excited CF levels Hanazawa \etal \cite{HANAZ85} 
have introduced a second Kondo temperature at high temperature 
$T_{\rm K}^{\rm h}=\sqrt[3] {T_{\rm K}\Delta_{\rm CF}^{(1)}
\Delta_{\rm CF}^{(2)}}$. With the 
assumption that $\Delta_{\rm CF}^{(1)}$ and $\Delta_{\rm 
CF}^{(2)}$ are hardly changed at low 
pressure (i.~e.~$P<4$~GPa), the $T_{\rm K}^{\rm h}$ values can be 
calculated if $T_{\rm K}$ is known. For some \cca compounds
$T_{\rm K}$ has been determined \cite{LOEHN96a}. 
To transform the $T_{\rm K}(V(x))$--dependence in a $T_{\rm 
K}(P)$--dependence the relative unit--cell volumes $V(x)/V_0$, with 
$V_0$ the unit--cell volume of \ccfa at ambient pressure, have to be 
transformed into the corresponding pressure values. With the 
Murnaghan equation of state (EOS) \cite{MURNA44} a $T_{\rm K}(P)$ 
relation, applicable for CeCu$_5$Au, can be deduced (using
$B_0=110$~GPa and $B'_0=4$). This (linear) function then yields the 
$T_{\rm K}^{\rm h}(P)$ dependence. The $T_{\rm K}^{\rm h}(P)$ values 
are practically identical to $T_{\rm max}^{\rm high}(P)$ in the 
pressure range $2<P<3$~GPa. At lower pressure the agreement is not 
so good which might be related to the presence of magnetic order. 

%############################################################################
%                                                                           #
%                   FIGURE 3                                                #
%                                                                           #
%############################################################################
\begin{figure}
\epsfxsize=100mm \epsfbox{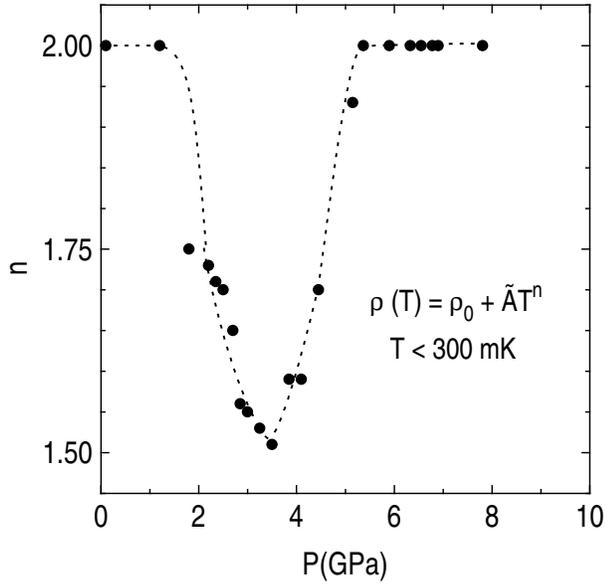}
\vspace*{-6cm}
\caption{The exponent $n$ used in the power law
of eq.~(1) to describe the $\rho(T)$ data of \ccfa below
$T=0.3$~K. Around $P=3.5$~GPa a clear deviation from a FL ($n=2$)
behaviour is observed. The dashed line is a guide to the eye.}
\label{nflccfa}
\end{figure}

Applying pressure to stoichiometric and single crystalline \ccfa
offers the possibility to study the low temperature
properties close to the magnetic instability and to compare them
with the \cca solid solution (with $x\rightarrow x_c$).
The electrical resistivity of \ccfa below 0.3~K can be described with

\begin{equation}
\rho =\rho_0 + \tilde{A}T^n
\label{freen}
\end{equation}
\noindent
at all pressures. The exponent $n$ and the coefficient $\tilde{A}$ 
are fitting parameters. The only "constraint" to the fit was the 
fixation of the upper temperature limit $T_{\rm A}=0.3$~K. It is a 
compromise between an as narrow as possible temperature interval 
(30~mK$<T<300$~mK) and the reliability of the deduced parameters, 
i.~e.~$n$ and $\tilde{A}$. Figure~\ref{nflccfa} illustrates how the 
deviation from a FL description ($n=2$) evolves with pressure. Below 
1.2~GPa $n=2$ is consistent with residual electron--magnon 
scattering in a magnetic system \cite{COQBL77}. Then, at 1.8~GPa, 
$n$ suddenly attains a value of 1.75 and decreases further as 
pressure increases. At 3.5~GPa $n=1.51$ is reached, close to the 
critical value $n=3/2$, predicted by theory \cite{MORIY95}. 
Increasing pressure further, leads to a higher $n$ value which 
finally reaches $n=2$, well inside the non--magnetic region ($P\geq 
5.37$~GPa) and comparable to CeCu$_6$ \cite{YOMO88}. The minimum in 
$n$ vs $P$ is not an artefact of the limited temperature interval in 
the fitting procedure as the fits for various $T_{\rm A}$ values (up 
to 0.6~K) showed always a minimum in $n(P)$ around 3.5~GPa, where 
$n$ is the smaller ($n=1.2$) the higher the $T_{\rm A}$ limit 
(0.6~K) was chosen. In the high pressure non--magnetic region 
however, the FL value $n=2$ was found in temperature intervals which 
became enlarged with pressure. 
%, i.~e.~$T_{\rm A}$ increased with pressure. 

This behaviour immediately raises the question whether these results
can be compared to the observations in \cca with different
Au concentrations. It is clear that the unit--cell volume variation
is a crucial parameter. Thus, a correspondence of the $x$ values in
\cca to the pressure values in \ccfa is needed. Using the
relation $V(x)=420.225~$\AA$^3 + 13.988x$, deduced from x--ray data 
\cite{SCHLA93} and e.~g.~the Murnaghan EOS, a 
relation between $x$ and $P$ can be obtained. For CeCu$_5$Au, this results in
a bulk modulus $B_0=110$~GPa (with $B'_0=4$) leading to
the correspondence $x=0.5\Leftrightarrow P=1.8$~GPa, 
$x=0.1\Leftrightarrow P=3.4$~GPa, and $x=0\Leftrightarrow 
P=3.85$~GPa which should be taken as a guide rather than as 
a strict prediction. 
%############################################################################
%                                                                           #
%                   FIGURE 4                                                #
%                                                                           #
%############################################################################

\begin{figure}
\epsfxsize=100mm \epsfbox{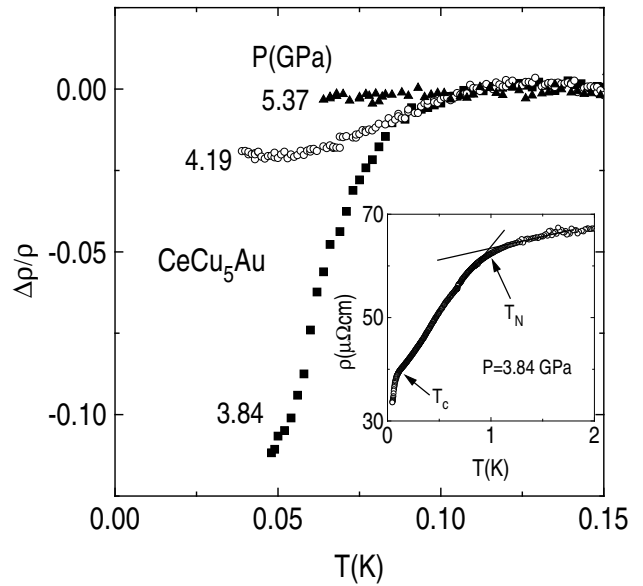}
\vspace*{-5.5cm}
\caption{Electrical resistivity $\rho(T)$ of \ccfa normalized 
at $T=0.1$~K at pressures close to the magnetic instability. The 
strong drop in $\rho(T)$ at $T=0.1$~K and $P=3.84$~GPa is 
interpreted as the entrance into a (probably) superconducting phase. 
Inset: The anomaly in $\rho(T)$ at $\approx~1$~K is interpreted as a 
sign of magnetic order.}
\label{ccfasupra}
\end{figure}

Now the peculiar low temperature behaviour of the $\rho(T)$ curves 
recorded close to the magnetic instability are of particular 
interest (Fig.~\ref{ccfasupra}). At $P=3.84$~GPa the entrance into a 
magnetically ordered phase at $T_{\rm N}\approx 1$~K is still 
visible (inset Fig.~\ref{ccfasupra}) but at $T=0.1$~K, $\rho(T)$ 
drops suddenly more than 10\%. This effect can be suppressed if a 
small magnetic field ($\vec{B}\parallel\vec{c}$, $B=0.2$~T) is 
applied. Traces of this transition are also present at $P=4.19$~GPa, 
where the resistivity starts to decrease, but not as strong as at the 
preceding pressure. No signs of a magnetically ordered phase were 
found. Hence, if magnetism and superconductivity coexist, it occurs 
in a very narrow pressure range. Well above this pressure, no 
anomalies in the low temperature part of $\rho(T)$ were found (see
curve at $P=5.37$~GPa in Fig.~\ref{ccfasupra}). 

If the new phase should be found unequivocally to be a 
superconducting phase, the measurements show that low residual 
resistivity is not an important ingredient for superconductivity for 
CeCu$_5$Au. Just before $\rho(T)$ of \ccfa starts to decrease, the 
resistivity is close to 40~$\mu\Omega$cm (at 0.1~K and 
$P~=~3.84$~GPa). Furthermore, as in the other pressure--induced HF 
superconductors, superconductivity then would emerge in the vicinity 
of the magnetic instability. Therefore, it is an intriguing 
possibility that AFM spin fluctuations may provide the attractive 
interaction between quasi--particles which is required to form 
Cooper--pairs \cite{MATHU98}. However, additional experiments are necessary to 
clarify this point. 

%###########################################################################
%                                                                          #
%                      Conclusion                                          #
%                                                                          #
%###########################################################################

\begin{flushleft}
{\bf 3. Conclusion}
\end{flushleft}
The electrical resistivity measurements on single crystalline \ccfa 
showed that the long--range magnetic order is suppressed at  $P_{\rm 
c}=4.1(3)$~GPa. Close to the magnetic instability the electrical 
resistivity (below $T=0.3$~K) deviates from a Fermi--liquid 
behaviour. At $P=3.84$~GPa the pronounced drop in $\rho(T)$ at 
$T_{\rm c}=0.1$~K  might point to the existence of a superconducting 
phase. At this pressure an AFM order is still present ($T_{\rm 
N}\approx 1$~K), leading to the possibility to study the coexistence 
of AFM order and superconductivity as well as NFL behaviour close to 
a quantum critical point. 

\begin{flushleft}
{\bf 4. Acknowledgments}
\end{flushleft}
This work was partially supported by the Swiss National Science Foundation.

\vspace*{0.5cm}
\begin{flushleft}
{\bf References}
\end{flushleft}


\vspace*{-2cm}

\begin{references}
\bibitem{MATHU98} N. D. Mathur, F. M. Grosche, S. R. Julian, I. R. Walker, R. K. W. Haselwimmer,
and G. G. Lonzarich, Nature {\bf 394}, 39 (1998).

\bibitem{WILHE99} H. Wilhelm, K. Alami-Yadri, B. Revaz, and D. Jaccard,
Phys. Rev. B {\bf 59}, 3651 (1999).

\bibitem{LOEHN94} H. v. L\"ohneysen, T. Pietrus, G. Portisch, H. G. Schlager, A. Schr\"oder,
M. Sieck, and T. Trappmann, Phys. Rev. Lett. {\bf 72}, 3262 (1994).

\bibitem{HERTZ76} J. A. Hertz, Phys. Rev. B {\bf 14}, 1165 (1976).

\bibitem{MILLI93} A. J. Millis, Phys. Rev. B {\bf 48}, 7183 (1993).

\bibitem{MORIY95} T. Moriya and T. Takimoto, J. Phys. Soc. Jpn. {\bf 64},
960 (1995).

\bibitem{HUXLE96} A. Huxley, S. Kambe, C. Pfleiderer, H. Suderow,
C. Thessieu, A. Buzdin, J. Flouquet, L. Gl\'emot, I. Fomin, 
and J.-P. Brison, J. Phys. Soc. Jpn. {\bf 65}, Suppl. B, 1 (1996).

\bibitem{GEGEN99} P. Gegenwart, F. Kromer, M. Lang, G. Sparn, C. Geibel,
and F. Steglich, Phys. Rev. Lett. {\bf 82}, 1293 (1999).

\bibitem{LOEHN98} H. v. L\"ohneysen, A. Neubert, A. Schr\"oder, O. Stockert, U. Tutsch,
M. Loewenhaupt, A. Rosch, and P. W\"olfle,
Eur. Phys. J. B {\bf 5}, 447 (1998).

\bibitem{COLEM99} P. Coleman, Physica B {\bf 259-261}, 353 (1999).

\bibitem{ROSCH99} A. Rosch, Phys. Rev. Lett. {\bf 82}, 4280 (1999).

%\bibitem{GEGEN98} P. Gegenwart, C. Langhammer, C. Geibel, R. Helfrich, M. Lang,
%G. Sparn, F. Steglich, R. Horn, L. Donnevert, A. Link, and W. Assmus,
%Phys. Rev. Lett. {\bf 81}, 1501 (1998).

\bibitem{ROSCH97} A. Rosch, A. Schr\"oder, O. Stockert, and H. v. L\"ohneysen,
Phys. Rev. Lett. {\bf 79}, 159 (1997).

\bibitem{STOCK98} O. Stockert, H. v. L\"ohneysen, A. Rosch, N. Pyka,
and M. Loewenhaupt, Phys. Rev. Lett. {\bf 80}, 5627 (1998).

\bibitem{PIETR95} T. Pietrus, B. Bogenberger, S. Mock, M. Sieck, and H. v. L\"ohneysen, 
Physica B {\bf 206\&207}, 317 (1995).

\bibitem{SCHLA93} H. G. Schlager, A. Schr\"oder, M. Welsch, and H. v. L\"ohneysen,
J. Low Temp. Phys. {\bf 90}, 181 (1993).

\bibitem{STEWA84} G. R. Stewart, Z. Fisk, amd M. S. Wire, Phys. Rev. B {\bf 30}, 482 (1984);
A. Amato, D. Jaccard, J. Flouquet, F. Lapierre, J. L. Tholence, R. A. Fisher, S. E. Lacy,
J. A. Olsen, and N. E. Phillips, J. Low Temp. Phys. {\bf 68}, 371 (1987).

\bibitem{STOCK99} O. Stockert, A. Schr\"oder, H. v. L\"ohneysen, N. Pyka, E. Garcia-Matres,
R. v. d. Kamp, S. Wezel, and M. Loewenhaupt, Physica B {\bf 259-261}, 383 (1999).

\bibitem{PASCH94} C. Paschke, C. Speck, G. Portisch, and H. v. L\"ohneysen,
J. Low Temp. Phys. {\bf 97}, 229 (1994).

\bibitem{WILHE99d} H. Wilhelm, S. Raymond, D. Jaccard, O. Stockert, and H. v. L\"ohneysen,
to be published.

\bibitem{STROK93} B. Stroka, A. Schr\"oder, T. Trappmann, H. v. L\"ohneysen, M. Loewenhaupt,
and A. Severing, Z. Phys. B {\bf 90}, 155 (1993).

\bibitem{HANAZ85} K. Hanazawa, K. Yamada, and K. Yoshida, J. Magn. Magn. Mater. {\bf 47\&48}, 357 (1985).

\bibitem{LOEHN96a} H. v. L\"ohneysen, M. Sieck, O. Stockert, and M. Waffenschmidt,
Physica B {\bf 223\&224}, 471 (1996).

\bibitem{MURNA44} F. D. Murnaghan, Proc. Natl. Acad. Sci. USA {\bf 30}, 244 (1944).

\bibitem{YOMO88} S. Yomo, L. Gao, R. L. Meng, P. H. Hor, C. W. Chu, and J. Susaki,
J. Magn. Magn. Mater. {\bf 76\&77}, 257 (1988).

\bibitem{COQBL77} B. Coqblin, in {\it The Electronic Structure of Rare--Earth Metals
and Alloys: The Magnetic Heavy Rare--Earths}, edited by B. Coqblin, Academic Press 1977.

%\bibitem{LOEHN96b} H. v. L\"ohneysen, J. Magn. Magn. Mater. {\bf 157/158}, 601 (1996).
%\bibitem{NEUBE97} A. Neubert, T. Pietrus, O. Stockert, H. v. L\"ohneysen, A.
%Rosch, and P. W\"olfle, Physica B {\bf 230-232}, 587 (1997).

%\bibitem{LOEHN98a} H. v. L\"ohneysen, S. Mock, A. Neubert, T. Pietrus,
%A. Rosch, A. Schr\"oder, O. Stockert, and U. Tutsch,
%J. Magn. Magn. Mat. {\bf 177-181}, 12 (1998).

\end{references}
\end{document}